\documentclass[nofootinbib,twocolumn,showpacs,preprintnumbers,pre,aps]{revtex4-1}

\usepackage[dvipdfmx]{graphicx}
\usepackage{bm}
\usepackage{amsmath}
\usepackage{amssymb}
\usepackage{color}

\begin{document}

\title{Nonreciprocality of a micromachine driven by a catalytic chemical reaction}

\author{Kento Yasuda}\email{Present address: Research Institute for Mathematical Sciences, Kyoto University, Kyoto 606-8502, Japan}

\author{Shigeyuki Komura}\email{komura@tmu.ac.jp}

\affiliation{
Department of Chemistry, Graduate School of Science,
Tokyo Metropolitan University, Tokyo 192-0397, Japan}


\begin{abstract}
We propose a model that describes cyclic state transitions of a micromachine driven by a catalytic chemical reaction.
We consider a mechano-chemical coupling of variables representing the degree of a chemical reaction and 
the internal state of a micromachine.
The total free energy consists of a tilted periodic potential and a mechano-chemical coupling energy.
We assume that the reaction variable obeys a deterministic stepwise dynamics characterized by two 
typical time scales, i.e., the mean first passage time and the mean first transition path time.
To estimate the functionality of a micromachine, we focus on the quantity called ``nonreciprocality" 
and further discuss its dependence on the properties of catalytic reaction.
For example, we show that the nonreciprocality is proportional to the square of the mean first transition 
path time.
The explicit calculation of the two time scales within the decoupling approximation model reveals that the 
nonreciprocality is inversely proportional to the square of the energy barrier of catalytic reaction.
\end{abstract}

\maketitle

\section{Introduction}
\label{intro}

In recent years, physics of micromachines such as bacteria, motor proteins, and artificial molecular machines 
has been intensively studied~\cite{Toyabe15,Brown20}.
Generally, a micromachine can be defined as a small object that extracts energy from chemical substances in 
the system and further exhibits mechanical functions.
The interplay between the structural dynamics of such a small object and the associated chemical reaction is 
crucial for the operation of a micromachine~\cite{Togashi10,Mugnai20}.
Owing to the developments in nonequilibrium statistical mechanics and experimental techniques,
various researches have been conducted to reveal the energetics of a single micromachine.
For example, energy efficiencies of F$_1$-ATPase and kinesin motors have been experimentally 
measured by using the Harada-Sasa relation~\cite{Harada05,Toyabe10,Ariga18}.

Furthermore, attention has been paid to the dynamics of micromachines.
For instance, several works reported that diffusion coefficients of metabolic enzymes increase due to 
enzymatic reactions~\cite{Muddana10, Golestanian15, Ghosh21}.
Although various possible scenarios have been proposed such as self-thermophoresis, stochastic swimming,
or collective heating, the main physical mechanism for the enhanced diffusion is not yet specified~\cite{Golestanian15}. 
Moreover, the experiment by Jee \textit{et al.}\ showed that metabolic enzymes can move in a directional 
manner in the presence of catalytic reactions~\cite{Jee18}. 
Although such a swimming behavior can be explained by a specific theoretical model~\cite{Sakaue10}, 
a more fundamental understanding concerning the interplay between the dynamics and function of a micromachine 
is necessary.

\begin{figure}[tb]
\begin{center}
\includegraphics[scale=0.6]{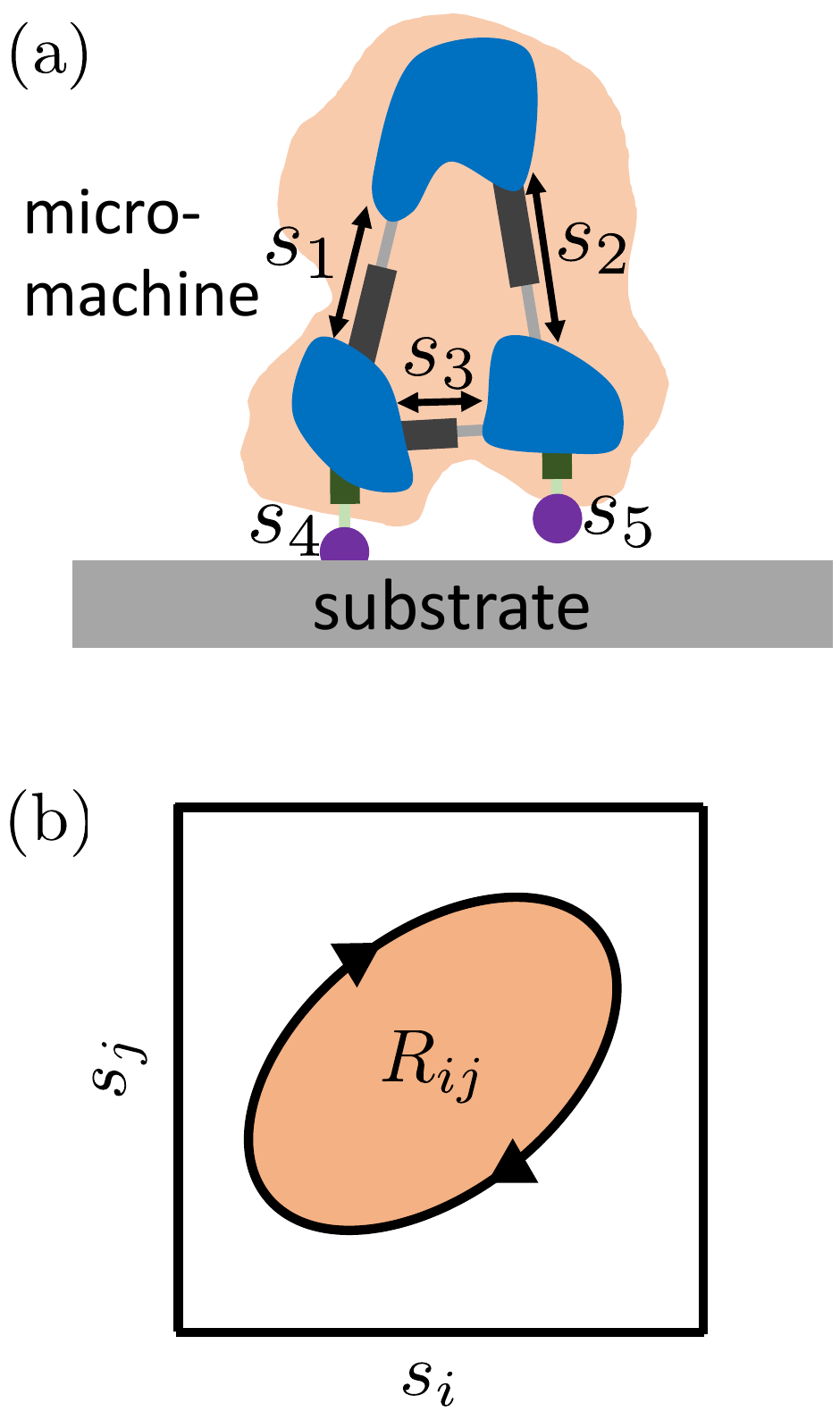}
\end{center}
\caption{
(a) Schematic picture of a micromachine characterized by the conformational state variables $s_1$, $s_2$, and $s_3$.
Moreover, the adhesion between the domains and the substrate is described by the variables 
$s_4$ and $s_5$.
(b) The state transition of a micromachine is represented by a trajectory in the state space spanned by the
variables $s_i$ and $s_j$. 
The nonreciprocality $R_{ij}$ [see Eq.~(\ref{Cyclon})] represents the area enclosed by the trajectory.}
\label{Fig:mod}
\end{figure}

Biological functions of a micromachine is intimately related to the transitions between different internal states.
As depicted in Fig.~\ref{Fig:mod}(a), one can use time-dependent state variables $s_i(t)$ 
($i=1,2,3,\cdots$) to characterize such as conformational structure or adhesion state when a micromachine 
is interacting with a substrate.
The state variables $s_i(t)$ change dynamically when a micromachine catalyzes a chemical reaction of substrate 
molecules.
As long as it acts as a catalyst, however, the internal state should return to the initial state after 
one cycle of reaction.
Hence, $s_i(t)$ should change periodically in time as the chemical reaction proceeds repeatedly.

In overdamped systems, such a cyclic change of internal state is related to biological functions of 
a micromachine. 
For example, microswimmers in a viscous fluid have been investigated by using specific 
models such as connected spheres~\cite{Golestanian08, Avron05} or a spherical 
squirmer~\cite{Laugabook}.
It was shown that the average swimming velocity is proportional to the closed  loop area in the 
corresponding deformation space. 
In order to extend this concept and to generally characterize the functionality of a generic micromachine, 
we focus on the following quantity 
\begin{align}
R_{ij}&=\int_{0}^{\tau_\mathrm{c}}dt\,\dot s_i  s_j,
\label{Cyclon}
\end{align}
where $\dot{s}_{i}=ds_{i}/dt$ and $\tau_\mathrm{c}$ is the period of the cycle.
In this paper, we shall call $R_{ij}$ as ``nonreciprocality" representing the area enclosed 
by a trajectory in the state space, as shown in Fig.~\ref{Fig:mod}(b). 

For a three-sphere microswimmer~\cite{Golestanian08}, $s_1$ and $s_2$ correspond to the lengths of 
the two arms, and its average swimming velocity is directly proportional to the nonreciprocality, i.e., 
$V\sim R_{12}$.
This relation was also discussed in the experimental realization of a three-sphere microswimmer~\cite{Grosjean16}.
Such a relation holds not only for deterministic microswimmers, but also for stochastic microswimmers~\cite{Sou19,Sou21}.
The nonreciprocality $R_{ij}$ is also relevant to crawling motions of a cell on a substrate~\cite{Tarama18,Leoni17}.
Within a gauge theory, the average velocity of a deformable body is given by $V=\sum_{i,j}W_{ij}R_{ij}$, where 
$W_{ij}$ is a weighting tensor that connects the velocity and the nonreciprocality~\cite{Shapere89}.
It should be noted that the nonreciprocality $R_{ij}$ is a universal quantity that does not depend 
on specific self-propulsion models. 
Moreover, $R_{ij}$ quantifies how much a micromachine breaks the detailed balance that needs to 
be satisfied in thermal equilibrium.

Although the functionality of a micromachine can be quantified by $R_{ij}$, it is important to 
clarify how the nonreciprocality is regulated within a micromachine. 
Considering a micromachine that utilizes a catalytic chemical reaction, we investigate 
the relationship between the properties of the reaction and the nonreciprocality $R_{ij}$.
Our main purpose is to propose a minimum model of a micromachine undergoing cyclic state 
transitions which are driven by repeated catalytic reactions.
Hence, in addition to the state variables $s_i(t)$, we introduce another variable $\theta(t)$ to 
describe the degree of a catalytic chemical reaction.
These variables are related to each other through the mechano-chemical coupling mechanism.

We employ Onsager's phenomenological equations for the time evolutions of $s_i(t)$ and $\theta(t)$.
In order to solve the equations analytically, we consider the weak coupling limit and assume that the 
dynamics of $\theta(t)$ is described by a step function characterized by two characteristic time scales, 
i.e., the mean first passage time $\tau_{\rm p}$ and the mean first transition path time $\tau_{\rm t}$.
Solving the equations for the state variables $s_i$, we analytically obtain the nonreciprocality $R_{12}$ 
as a function of the above two time scales.
Furthermore, we obtain analytical expressions of $\tau_{\rm p}$ and $\tau_{\rm t}$ within the 
decoupling approximation, and relate them to the properties of catalytic reaction.
Combining these results, we show that the nonreciprocality is inversely proportional to the square 
of the energy barrier of catalytic reaction.

In the next section, we explain the model of a micromachine driven by a catalytic chemical reaction.
In Sec.~\ref{state}, we argue the dynamics of the state variables $s_{i}$.
In Sec.~\ref{cyclone}, we calculate the nonreciprocality $R_{12}$ analytically.
In Sec.~\ref{time}, we obtain the mean first passage time $\tau_{\rm p}$ and the mean first transition 
path time $\tau_{\rm t}$.
Finally, a summary of our work and some further comments are given in Sec.~\ref{diss}.

\section{Model}
\label{model}

\subsection{Catalytic chemical reaction}

Consider a system which contains one enzyme molecule (E) that acts as a micromachine, 
$n_\mathrm{S}$ substrate molecules (S), and $n_\mathrm{P}$ product molecules (P).
The enzyme molecule plays the role of a catalyst and the corresponding chemical reaction is written 
as~\cite{Dillbook} 
\begin{align}
\mathrm{S}+\mathrm{E}\rightleftarrows \mathrm{ES}\to \mathrm{P}+\mathrm{E}
\end{align}
where ES indicates a complex molecule.
The reaction rate $\dot n_\mathrm{P}$ is often analyzed by the Michaelis-Menten equation~\cite{Dillbook}.
Although the above catalytic chemical reaction is relevant to the present study, our purpose is to adopt 
the simplest model for such a chemical reaction and not to reproduce it.

The extent of a catalytic reaction is commonly described by the number of product molecules $n_\mathrm{P}$.
However, since our purpose is to investigate a single molecular reaction process, we introduce a reaction 
variable $\theta(t)$ to quantify the extent of catalytic reaction.
Unlike the quantity $n_\mathrm{P}$, the reaction variable $\theta$ is a continuous number and increases $2\pi$ 
for each reaction.
Under this assumption, $\theta$ represents the reaction phase of a periodic catalytic reaction.

\begin{figure}[tb]
\begin{center}
\includegraphics[scale=0.6]{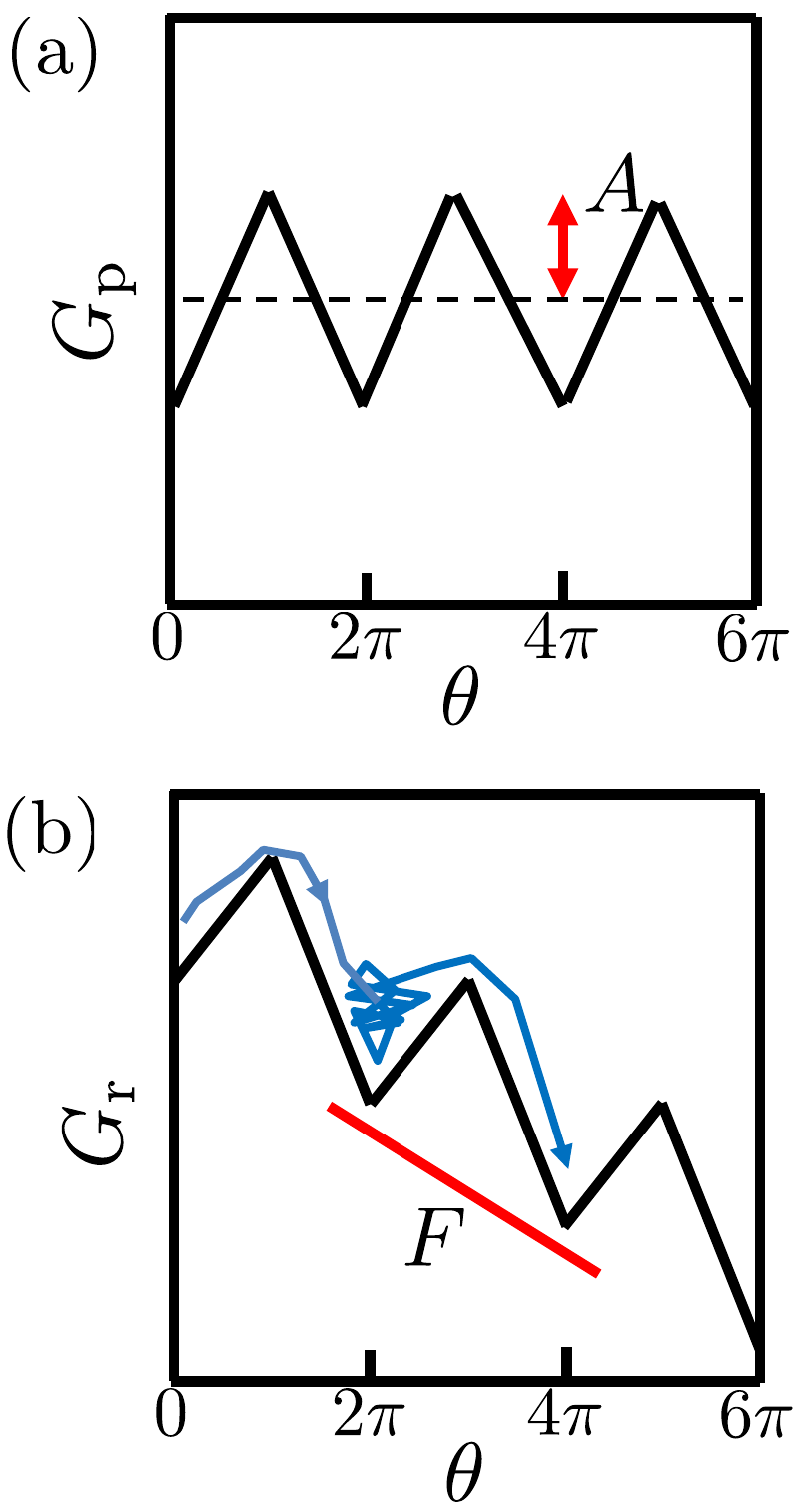}
\end{center}
\caption{
(a) The periodic potential $G_\mathrm{p}$ as a function of the catalytic reaction variable $\theta$
with a period of $2\pi$.
Here we show a linear function given by the Eq.~(\ref{Gp}) with an energy barrier $A$. 
(b) The tilted periodic potential $G_\mathrm{r}$ as a function of the reaction variable $\theta$. 
As shown in Eq.~(\ref{Gt}), $G_\mathrm{r}$ consists of the periodic part $G_\mathrm{p}$ in (a) and the 
linear part $-F\theta$, where $F$ is the nonequilibrium force.
A possible trajectory of $\theta$ is shown by the blue (gray) arrow.
The value of $\theta$ fluctuates around the minimum of the potential and the transition to the next minimum 
takes place occasionally.
}
\label{Fig:pot}
\end{figure}

According to the Kramers theory, the free energy $G_{\mathrm{r}}$ describing a chemical reaction is given 
by a tilted periodic potential~\cite{Hanggi90} 
\begin{align}
G_{\mathrm{r}}(\theta)=G_\mathrm{p}(\theta)-F\theta,
\label{Gt}
\end{align}
where $G_\mathrm{p}$ is a periodic potential with a period of $2\pi$, i.e., 
$G_\mathrm{p}(\theta+2\pi)=G_\mathrm{p}(\theta)$, as shown schematically in Fig.~\ref{Fig:pot}(a).
This is because $\theta$ increases by $2\pi$ for one cycle of chemical reaction and should experience
the same potential.
We also require that $G_\mathrm{p}$ takes minimum values at $\theta = 2n\pi$ ($n$ being an integer)
because the chemical states should be stable before and after the catalytic reaction.
As shown in Fig.~\ref{Fig:pot}(a), the amplitude of $G_\mathrm{p}$, denoted by $A$, represents the 
energy barrier in the chemical reaction and is regarded as the activation energy.
The explicit form of $G_\mathrm{p}$ will be presented later in Eq.~(\ref{Gp}).

On the other hand, $F$ in Eq.~(\ref{Gt}) represents the chemical potential difference 
(such as between ATP and ADP molecules) that drives catalytic reaction.
Physically, it corresponds to a nonequilibrium force even though $F$ has the dimension of energy.
The system is in chemical equilibrium when $F=0$, whereas it is in out-of-equilibrium situation when $F\ne 0$. 
In this paper, we shall consider the case of $F>0$.
With the added nonequilibrium force $F$, the free energy $G_{\mathrm{r}}$ for catalytic reaction
becomes a tilted periodic potential as schematically shown in Fig.~\ref{Fig:pot}(b).

\subsection{Mechano-chemical coupling}

Next, we introduce the state variables $s_i(t)$ ($i=1,2,3,\cdots$) characterizing the conformation 
of a micromachine.
As shown in Fig.~\ref{Fig:mod}(a), examples of the state variables are distances between the domains
in a micromachine or distances between the domains and the substrate (if it exists). 
In the molecular dynamics simulation of myosin V, for example, the protein structure is characterized by 
the relative distances between the three amino acids~\cite{Togashi10}.
In principle, there are a large number of degrees of freedom of a micromachine, and hence the number of 
the state variables $s_i$ can also be large.

Next, we explain the mechano-chemical coupling mechanism in our model.
We assume that each state variable $s_i$ experiences a harmonic potential, $K_i[s_i-\ell_i(\theta)]^2/2$, 
where $K_i$ is the coupling parameter and $\ell_i(\theta)$ is the natural state that is a function of the 
reaction variable $\theta$.
For a catalytic reaction, the internal state of a micromachine should return to the initial state after 
one cycle of reaction and the same process takes place repeatedly.
Hence, we consider that the natural state $\ell_i(\theta)$ changes also periodically and assume 
the simplest periodic form $\ell_i (\theta)=d_i \sin(\theta+\phi_i)$,
where $d_i$ is the amplitude and $\phi_i$ is the constant phase difference relative to the 
reaction phase $\theta$.
Under these assumptions, we consider the following mechano-chemical coupling energy $G_\mathrm{c}$
between 
$\theta$ and $s_i$:
\begin{align}
G_\mathrm{c}(\theta,\{s_i\})=\sum_i\frac{K_i}{2}\left[ s_i-d_i\sin(\theta+\phi_i)\right]^2.
\label{Gs}
\end{align}
Then the total free energy $G_{\mathrm{t}}$ in our model is simply given by 
\begin{align}
G_{\mathrm{t}}(\theta,\{s_i\})=G_\mathrm{r}(\theta)+G_\mathrm{c}(\theta,\{s_i\}).
\end{align}

\subsection{Dynamic equations}

For the time evolutions of $\theta$ and $s_i$, we employ the Onsager's phenomenological 
equations~\cite{Doibook}
\begin{align}
&\dot{\theta}=-M\frac{\partial G_{\mathrm{t}}}{\partial\theta}+\xi(t),
\label{Dt1}\\
&\dot{s_i}=-\sum_{j}\mu_{ij}\frac{\partial G_{\mathrm{t}}}{\partial s_j}+ \xi_i(t),
\label{Ds1}
\end{align}
where $M$ and $\mu_{ij}$ are the Onsager coefficients for $\theta$ and $s_i$, respectively.
These coefficients represent energy dissipation, and $\mu_{ij}$
is given, for example, by the inverse of the friction coefficient of a domain due to the surrounding viscous fluid.
Moreover, $\xi$ and $\xi_i$ represent thermal fluctuations which satisfy the fluctuation-dissipation theorem
\begin{align}
&\langle\xi(t)\rangle=0,\\
&\langle\xi(t)\xi(t')\rangle =2Mk_\mathrm{B}T\delta(t-t'),\\
&\langle\xi_i(t)\rangle=0,\\
&\langle \xi_i(t)\xi_j(t')\rangle =2 \mu_{ij}k_\mathrm{B}T\delta(t-t'),
\end{align}
where $k_{\rm B}$ is the Boltzmann constant and $T$ is the temperature.

In the above equations, the Onsager coefficients are assumed to be constant and thermal 
fluctuations are given by Gaussian white noise.
In the presence of a memory effect such as viscoelasticity, the Onsager coefficients depend
on time and thermal fluctuations are given by colored noise in the form of generalized 
fluctuation-dissipation relations.
In the absence of thermal fluctuations, the reaction variable $\theta$ does not evolve in time 
because of the energy barrier $A$ in the potential $G_\mathrm{r}$ [see Fig.~\ref{Fig:pot}(b)].
If thermal fluctuations are present, the value of $\theta$ fluctuates around the minimum of the 
potential and the transition to the next minimum takes place occasionally [blue (gray) trajectory
in Fig.~\ref{Fig:pot}(b)].
Hence, thermal fluctuations are necessary to drive time evolutions of $\theta$ and $s_i$ 
in our stochastic model.

Although our model is general, we make several simplifications in order to solve the 
coupled equations analytically.
First, we only take into account two degrees of freedom, i.e., $s_1$ and $s_2$.
Second, the mobility coefficients $\mu_{ij}$ is assumed to have the form 
$\mu_{11}=\mu_{22}=\mu$ and $\mu_{12}=\mu_{21}=0$. 
Third, the coupling free energy is symmetric between the two degrees of freedom, i.e., 
$K_1=K_2=K$ and $d_1=d_2=d$. 
Then Eqs.~(\ref{Dt1}) and (\ref{Ds1}) reduce to 
\begin{align}
\dot\theta&=-M[\partial_\theta G_\mathrm{r}(\theta)-Kd\cos(\theta+\phi_1)\delta_1\nonumber\\
&-Kd\cos(\theta+\phi_2)\delta_2]+\xi,
\label{Dt2}\\
\dot {\delta}_1&=-\gamma \delta_1-d\cos(\theta+\phi_1)\dot\theta+\xi_1,
\label{Ds2a}\\
\dot {\delta}_2&=-\gamma \delta_2-d\cos(\theta+\phi_2)\dot\theta+\xi_2,
\label{Ds2b}
\end{align}
where we have introduced $\delta_i=s_i-d\sin(\theta+\phi_i)$ and defined the 
relaxation rate $\gamma=\mu K$.

\begin{figure}[tb]
\begin{center}
\includegraphics[scale=0.3]{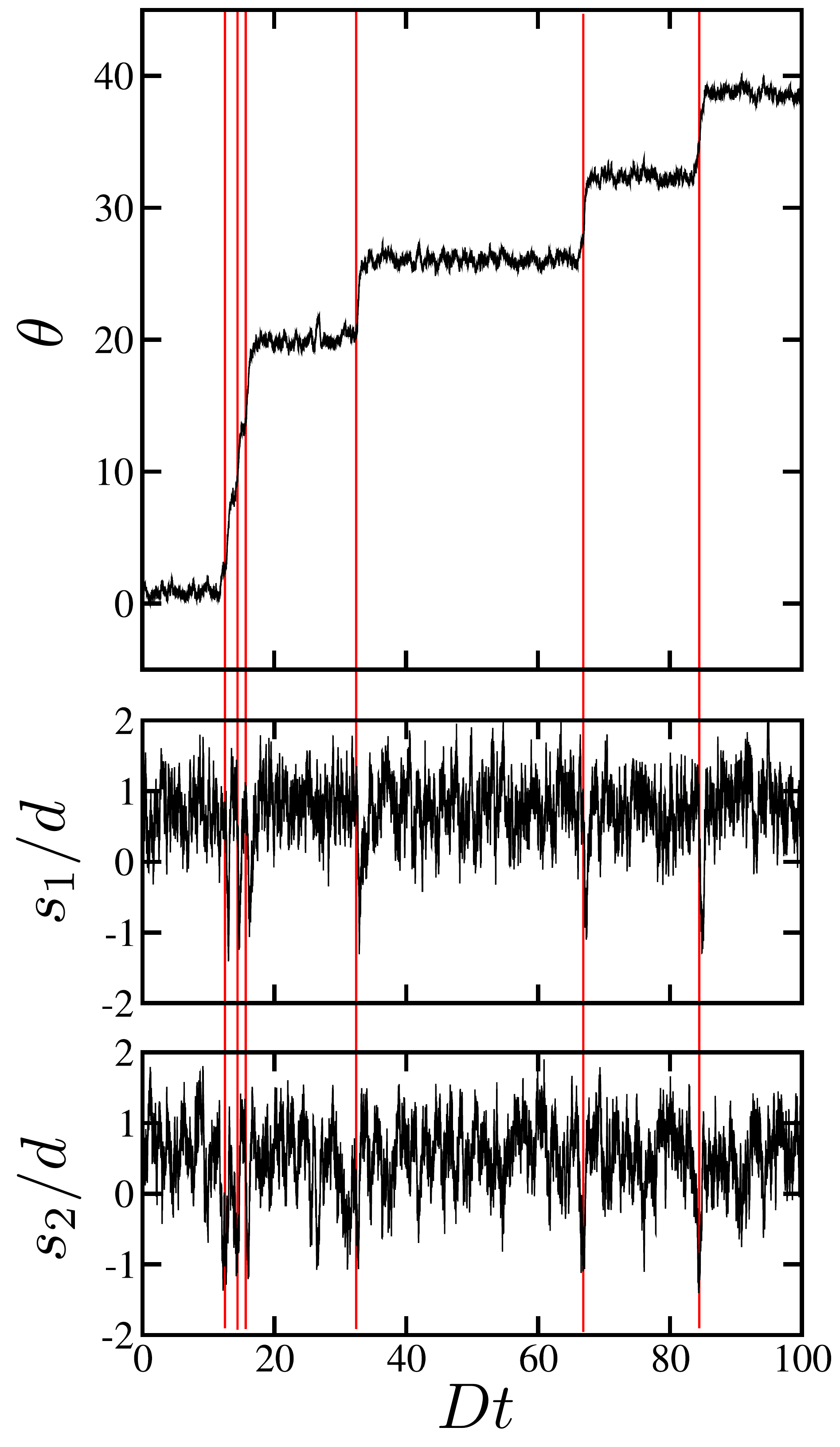}
\end{center}
\caption{
Time evolutions of $\theta$ (top), $s_1$ (middle), and $s_2$ (bottom) obtained by numerically solving
Eqs.~(\ref{Dt2})--(\ref {Ds2b}).
We use the dimensionless time by $Dt$, where $D=Mk_\mathrm{B}T$ has the dimension of inverse time.
The parameters are $a=A/(k_\mathrm{B}T)=10$, 
$f=F/(k_\mathrm{B}T)=8$, 
$Kd^2/(k_\mathrm{B}T)=10$, 
$\phi_1=0$, $\phi_2=\pi/2$, and $\mu/(d^2M)=1$. 
In this simulation, the periodic potential is taken to be $G_\mathrm{p}(\theta)=-A\cos\theta$ and is
different from the linear function in Eq.~(\ref{Gp}).
The red (gray) vertical lines indicate the moments when the catalytic chemical reactions take place.
}
\label{Fig:sim}
\end{figure}

As a demonstration of our model, we have numerically solved Eqs.~(\ref{Dt2})--(\ref {Ds2b}) by using the Euler method
and plotted in Fig.~\ref{Fig:sim} the time evolutions of $\theta$ (top), $s_1$ (middle), and 
$s_2$ (bottom) for certain parameters. 
We see that the reaction variable $\theta$ increases stochastically in a stepwise manner, 
whereas the state variables $s_1$ and $s_2$ undergo almost random fluctuations. 
Corresponding to the stepwise increase of $\theta$, both $s_1$ and $s_2$ tend to show 
peaks as indicated by the vertical red (gray) lines. 
Although the simulation result in Fig.~\ref{Fig:sim} demonstrates that the present minimum 
micromachine is indeed driven by thermal fluctuations, it is difficult to isolate 
the peaks of $s_{i}$ because they are almost comparable to the background fluctuations.
Moreover, the state transitions become very rare when $A\gg k_\mathrm{B}T$.
Therefore, instead of performing further simulations, we investigate our model analytically
by using the decoupling approximation.

\subsection{Decoupling approximation}

The decoupling approximation relies on two assumptions: (i) taking the weak coupling limit,
$K d^2 \ll A$, and (ii) assuming a deterministic dynamics for $\theta(t)$. 
Under the assumption (i), Eq.~(\ref{Dt2}) can be simplified to  
\begin{align}
\dot\theta=-M\partial_\theta G_\mathrm{r}(\theta)+\xi,
\label{Dt3}
\end{align}
where $\theta$ is now decoupled from $s_i$. 
If we were able to solve Eq.~(\ref{Dt3}) for $\theta$, we can further solve Eqs.~(\ref{Ds2a}) and (\ref{Ds2b}) 
for $\delta_i$ ($s_i$).
However, it is still difficult to solve Eq.~(\ref{Dt3}) analytically because $\theta$ is mostly trapped in 
the local minimum of $G_\mathrm{r}$ and thermal fluctuations are necessary to 
overcome the energy barrier $A$.

To tackle this problem, we further employ the assumption (ii) for $\theta(t)$.
As we shall explain in the next section, we assume that $\theta(t)$ is described by a 
deterministic function characterized by two time scales, i.e.,  
the mean first passage time $\tau_\mathrm{p}$ and the mean first transition 
path time $\tau_\mathrm{t}$ [see later Eq.~(\ref{Asst2}) and Fig.~\ref{Fig:asum}(b)].
Then we can first solve Eqs.~(\ref{Ds2a}) and (\ref{Ds2b}) analytically and obtain the explicit 
expression for $\delta_i$, as we show in Sec.~\ref{state}.
This result will be used to compute the nonreciprocality $R_{12}$ analytically in Sec.~\ref{cyclone}.
In Sec.~\ref{time}, on the other hand, the two characteristic time scales $\tau_\mathrm{p}$ and 
$\tau_\mathrm{t}$ will be separately calculated by using Eq.~(\ref{Dt3}) within the 
decoupling approximation.

\section{Dynamics of state variables}
\label{state}

\begin{figure}[tb]
\begin{center}
\includegraphics[scale=0.6]{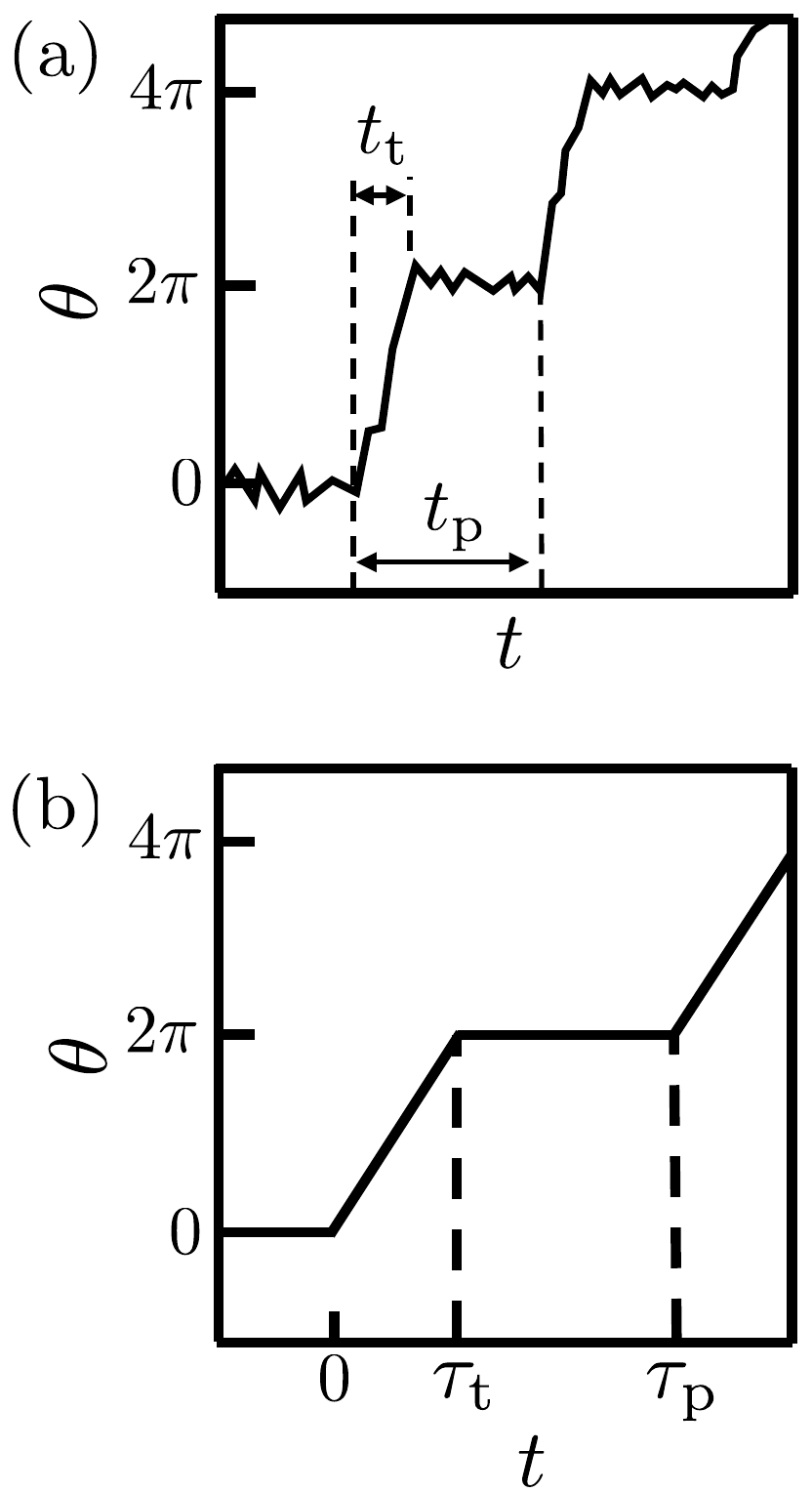}
\end{center}
\caption{
(a) Schematic example of a stochastic time evolution of the reaction variable $\theta$. 
Such a time evolution is characterized by the first transition path time $t_\mathrm{t}$ and 
the first passage time $t_\mathrm{p}$ which are both stochastic quantities.
(b) After averaging over these quantities, we obtain the average time evolution of $\theta$ 
as assumed in Eq.~(\ref{Asst2}).
Here $\tau_\mathrm{t}$ and $\tau_\mathrm{p}$ are the mean first transition path time and 
the mean first passage time, respectively.
The reaction variable increases linearly for $0 \le t<\tau_\mathrm{t}$ and remains constant 
for $\tau_\mathrm{t}\le t<\tau_\mathrm{p}$. 
The reaction phase $\theta(t)$ satisfies $\theta(t+\tau_\mathrm{p})=\theta(t) + 2\pi$.}
\label{Fig:asum}
\end{figure}

In this section, we discuss the dynamics of the state variables $\delta_1$ ($s_1$) and $\delta_2$
($s_2$) that obey Eqs.~(\ref{Ds2a}) and (\ref{Ds2b}), respectively.
To solve these equations, we make an assumption for the time dependence of $\theta$(t), 
as we mentioned in the previous section.
With a tilted periodic potential given by Eq.~(\ref{Gt}) and shown in Fig.~\ref{Fig:pot}(b),
the reaction variable $\theta$ changes stochastically and increases in a stepwise manner 
as we saw in the numerical simulation [see Fig.~\ref{Fig:sim} (top)] and also schematically 
depicted in Fig.~\ref{Fig:asum}(a).
Such a time evolution of $\theta(t)$ can be characterized by two characteristic time scales.
The first one is the ``first passage time'' $t_\mathrm{p}$ which is the time required to change from 
one local minimum to the neighboring lower local minimum~\cite{Hanggi90}.
The second one is the ``first transition path time'' $t_\mathrm{t}$ which is the time needed for the 
actual transition~\cite{Kim15,Chaudhury10,Hummer04}.
It should be noticed that both $t_\mathrm{p}$ and $t_\mathrm{t}$ are stochastic quantities.

\begin{figure*}[t]
\begin{center}
\includegraphics[scale=0.28]{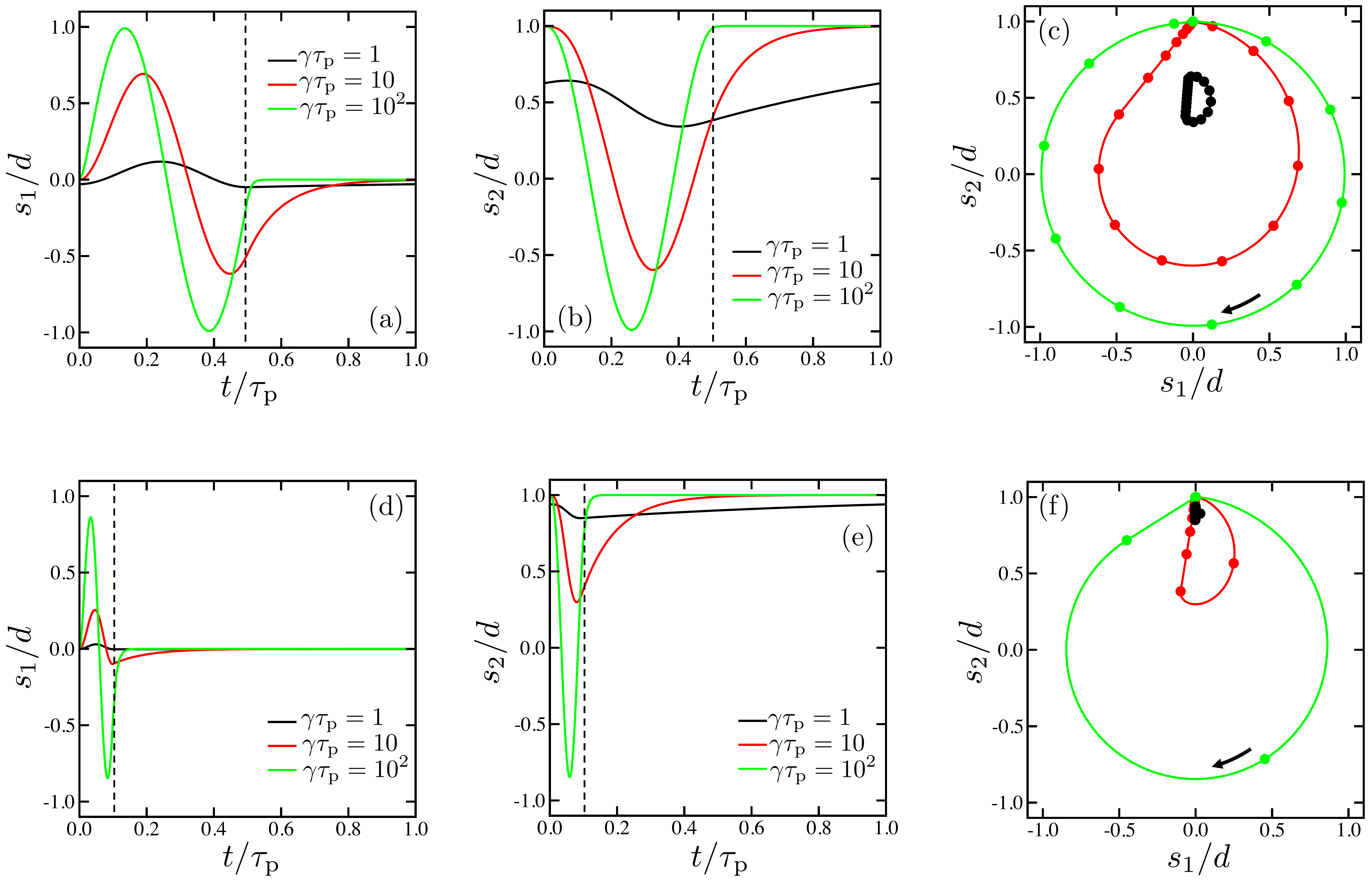}
\end{center}
\caption{
Time evolutions of (a) $s_1$ and (b) $s_2$ when $\phi_1=0$ and $\phi_2=\pi/2$. 
We set $\tau_\mathrm{t}/\tau_\mathrm{p}=0.5$ (shown by the dashed line)
and change $\gamma \tau_\mathrm{p}=1$ (black), $10$ (red or gray), and $10^2$ (green or light gray). 
(c) The trajectory of the state variables $s_1$ and $s_2$ shown in (a) and (b), respectively, over one cycle.
For each cycle, 20 equal time intervals are marked by the filled circles.
The black arrow indicates the direction of the state transition.
The enclosed area of the trajectory corresponds to the nonreciprocality $R_{12}$.
(d), (e), and (f) are the similar plots to (a), (b), and (c), respectively, when $\tau_\mathrm{t}/\tau_\mathrm{p}=0.1$
(shown by the dashed line).
}
\label{Fig:s}
\end{figure*}

To discuss the dynamics of $s_1$ and $s_2$, let us assume that $\theta$ can be represented by a 
deterministic stepwise function characterized by the ``mean first passage time'' $\tau_\mathrm{p}$ 
and the ``mean first transition path time'' $\tau_\mathrm{t}$ which are the averages of $t_\mathrm{p}$ 
and $t_\mathrm{t}$, respectively.
As depicted in Fig.~\ref{Fig:asum}(b), the assumed functional form of $\theta$ is 
\begin{align}
\theta(t)=
\begin{cases}
2\pi t/\tau_\mathrm{t} &{\rm for}~~~0 \le t < \tau_\mathrm{t}
\\
2\pi &{\rm for}~~~\tau_\mathrm{t} \le t<\tau_\mathrm{p}
\end{cases}.
\label{Asst2}
\end{align}
Furthermore, we require that $\theta$ increases by $2\pi$ after one cycle of catalytic reaction 
$\tau_\mathrm{p}$, i.e., $\theta(t+\tau_\mathrm{p})=\theta(t) + 2\pi$. 
The explicit expressions of $\tau_\mathrm{p}$ and $\tau_\mathrm{t}$ under the decoupling
approximation will be given in Sec.~\ref{time} where we focus on their dependencies on 
the energy barrier $A$ and the nonequilibrium force $F$.

Substituting Eq.~(\ref{Asst2}) into Eqs.~(\ref{Ds2a}) and (\ref{Ds2b}), we solve them in the absence of  
thermal noise, i.e., $\xi_{1}=\xi_{2}=0$  (see Appendix~\ref{appDs} for the details).
Then the stationary solution for $\delta_i$ ($i=1,2$) can be obtained as
\begin{align}
\delta_i(t)&=\frac{d}{(\gamma\tau_\mathrm t/2\pi)^2 + 1}\nonumber\\
&\times\left[-\frac{\gamma \tau_\mathrm t}{2\pi} 
\cos\left(\frac{2\pi t}{\tau_\mathrm t}+\phi_i\right)-\sin\left(\frac{2\pi t}{\tau_\mathrm t} +\phi_i\right)\right.\nonumber\\
&\left.+\frac{e^{\gamma \tau_\mathrm{p}}-e^{\gamma \tau_\mathrm{t}}}{e^{\gamma \tau_\mathrm{p}}-1}\left(\frac{\gamma \tau_\mathrm t}{2\pi} \cos\phi_i+
  \sin\phi_i\right)e^{-\gamma t}\right],
\label{Ss1}
\end{align}
for $0 \le t<\tau_\mathrm{t}$ and 
\begin{align}
\delta_i(t)&=-\frac{d}{(\gamma\tau_\mathrm t/2\pi)^2 + 1}\frac{e^{\gamma \tau_\mathrm{p}}(e^{\gamma \tau_\mathrm{t}}-1)}
{e^{\gamma \tau_\mathrm{p}}-1}\nonumber\\
&\times\left(\frac{\gamma \tau_\mathrm t}{2\pi} \cos\phi_i+ \sin\phi_i\right)e^{-\gamma t},
\label{Ss2}
\end{align}
for $\tau_\mathrm{t}\le t<\tau_\mathrm{p}$.

In Figs.~\ref{Fig:s}(a) and (b), we plot the time evolutions of $s_1$ and $s_2$, respectively, 
when $\phi_1=0$ and $\phi_2=\pi/2$. 
We set $\tau_\mathrm{t}/\tau_\mathrm{p}=0.5$ and change $\gamma \tau_\mathrm{p}=1$, $10$, and $10^2$.
Notice that $\gamma\tau_\mathrm{p}=\mu K\tau_\mathrm{p}$ is the dimensionless relaxation rate.
When the relaxation is fast, $\gamma\tau_\mathrm{p}>1$, the internal state of a micromachine can be sufficiently 
relaxed to the initial state within the reaction cycle $\tau_\mathrm{p}$.
When the relaxation is slow, $\gamma\tau_\mathrm{p}<1$, on the other hand, the next reaction starts before the 
internal state is fully relaxed.
The fast relaxation cases, $\gamma\tau_\mathrm{p}= 10$, $10^{2}$ adopted in Fig.~\ref{Fig:s} 
do not contradict with the weak coupling limit, $Kd^2 \ll A$, as we have discussed before.
These two conditions can be simultaneously satisfied when $\mu \gg d^2/(\tau_\mathrm{p}A)$, 
namely, when the Onsager coefficient $\mu$ is sufficiently large.

When $\gamma \tau_\mathrm{p}=10^2$ (green or light gray), the behaviors for $0\le t<\tau_\mathrm{t}$ are well described 
by sinusoidal functions $s_{1}/d=\sin(2\pi t/\tau_{\rm t})$ and $s_{2}/d=\cos(2\pi t/\tau_{\rm t})$.
In this fast relaxation case, we see a sufficiently large state change within a micromachine.
When $\gamma \tau_\mathrm{p}=1$ (black) or $10$ (red or gray), $s_i$ cannot follow the change in $\theta$ 
and the functionality of a micromachine is diminished.
In Fig.~\ref{Fig:s}(c), we plot the trajectories of $s_1$ and $s_2$ over one cycle of reaction for different 
values of $\gamma \tau_\mathrm{p}$.
For each cycle, 20 equal time intervals are marked by the filled circles.
As mentioned before, the enclosed area of each trajectory gives the nonreciprocality $R_{12}$.

In Figs.~\ref{Fig:s}(d), (e), and (f), we show the corresponding plots when $\tau_\mathrm{t}/\tau_\mathrm{p}=0.1$
(smaller $\tau_\mathrm{t}$).
When $\gamma \tau_\mathrm{p}=10^2$ (green or light gray) and $0\le t<\tau_\mathrm{t}$, both $s_1$ and $s_2$ are 
well described by the same sinusoidal functions as in Figs.~\ref{Fig:s}(a) and (b), respectively.
On the other hand, the black and red (gray) curves for $\gamma \tau_\mathrm{p}=1$ and $10$, respectively, 
deviate significantly from the green (light gray) curve for $\gamma \tau_\mathrm{p}=10^2$, and the magnitudes are 
significantly suppressed.
The reduced magnitudes can also be seen in Fig.~\ref{Fig:s}(f) where the areas enclosed by the black and red (gray)
lines are much smaller than that of the green (light gray) line.
This means that, for $\gamma \tau_\mathrm{p}=1$ and $10$, the nonreciprocality is further decreased as 
$\tau_\mathrm{t}/\tau_\mathrm{p}$ is made smaller.

\begin{figure}[tb]
\begin{center}
\includegraphics[scale=0.25]{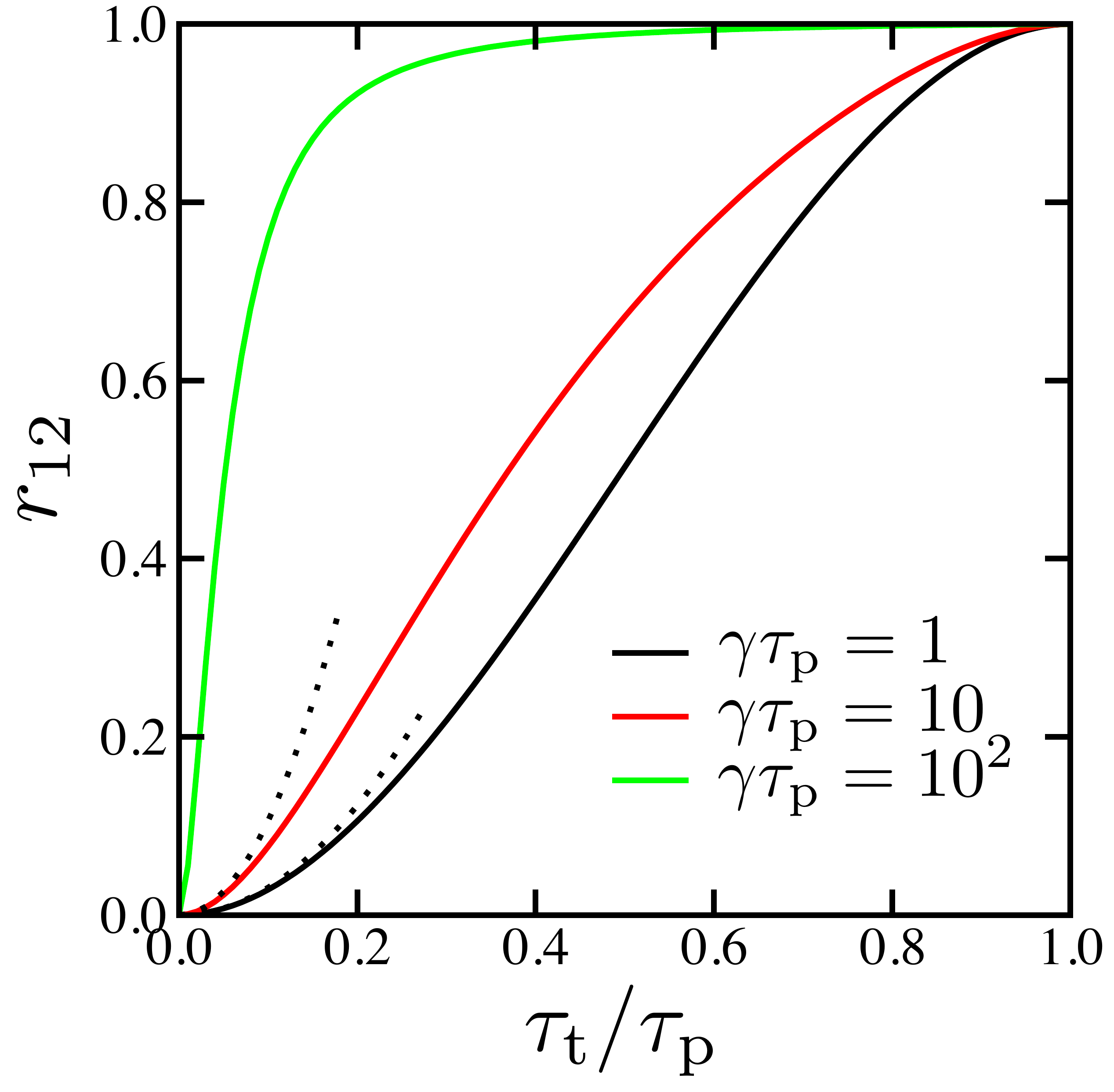}
\end{center}
\caption{
The dimensionless nonreciprocality $r_{12}$ (defined in the text) as a function of 
$\tau_\mathrm{t}/\tau_\mathrm{p}$ for $\gamma \tau_\mathrm{p}=1$ (black), $10$ (red or gray), and $10^2$ (green or light gray). 
The dotted lines are the asymptotic expression for $\tau_\mathrm{t}/\tau_\mathrm{p} \ll 1$ given by Eq.~(\ref{CA2}).
}
\label{Fig:C}
\end{figure}

\begin{figure}[tb]
\begin{center}
\includegraphics[scale=0.5]{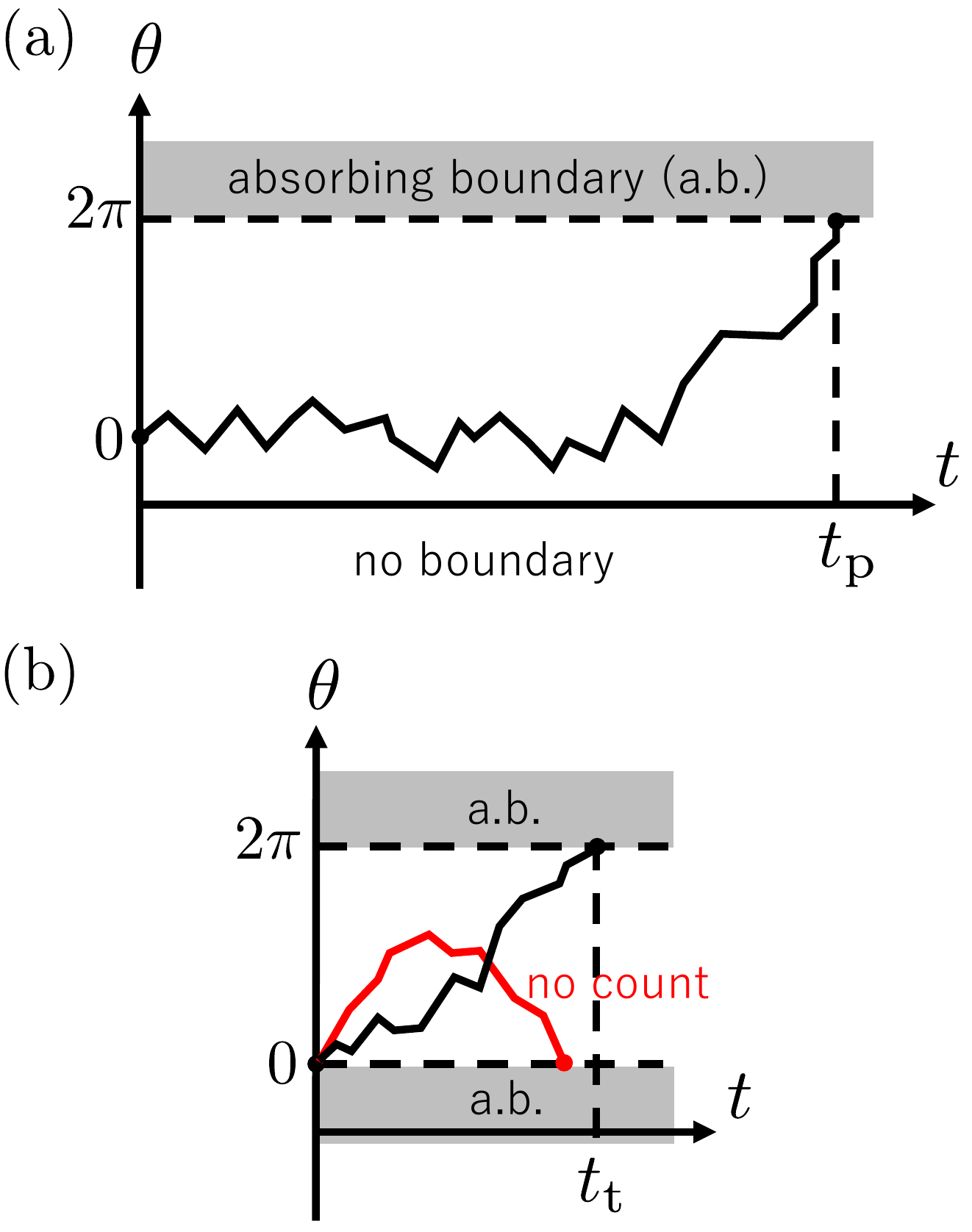}
\end{center}
\caption{
(a) Schematic description of the first passage time $t_\mathrm{p}$.
We consider a semi-infinite system with an absorbing boundary condition at $\theta=2\pi$,
and measure the time until a Brownian particle is absorbed at $\theta=2\pi$.
Here the state $\theta=0$ can be visited multiple times, which gives rise to a long waiting time.
(b) Schematic description of the first transition path time $t_\mathrm{t}$.
We consider a finite system with absorbing boundary conditions at $\theta=0$ and $2\pi$, and
measure the time until a Brownian particle is absorbed at $\theta=2\pi$.
When the particle is absorbed at $\theta=0$ [the red (gray) trajectory], such an event is not counted.
}
\label{Fig:TimeScale}
\end{figure}

\section{Nonreciprocality of a micromachine}
\label{cyclone}

We have mentioned in Introduction that the nonreciprocality defined in Eq.~(\ref{Cyclon}) provides us with a 
useful quantity to evaluate the functionality of a micromachine~\cite{Golestanian08, Shapere89,Sou19,Tarama18,Leoni17}.
Previously, the nonreciprocality was obtained along a deterministic state change when the 
period of deformation is constant.
However, this is not always possible when fluctuations are present~\cite{Ghanta17}.
For a stochastic micromachine, it is necessary either to take a long time limit or to estimate the statistical average
to estimate the nonreciprocality.
In the present model and analysis, on the other hand, one can calculate the nonreciprocality directly from 
Eq.~(\ref{Cyclon}) because we have assumed a deterministic dynamics for $\theta$ as in Eq.~(\ref{Asst2}).

With the use of Eqs.~(\ref{Ss1}) and (\ref{Ss2}), the nonreciprocality $R_{12}$ can be 
analytically obtained in terms of $\tau_\mathrm{p}$ and $\tau_\mathrm{t}$ as
\begin{align}
R_{12}&=\int_{0}^{\tau_\mathrm{p}}dt\,\dot s_1  s_2\nonumber\\
&=\frac{d^2\gamma^2 \tau_\mathrm{t}^2}{4\pi[(\gamma\tau_\mathrm{t}/2\pi)^2+1]}\sin(\phi_2-\phi_1)\nonumber\\
&\times\left[1+\frac{2\left[1+e^{\gamma \tau_\mathrm{p}}-e^{\gamma \tau_\mathrm{t}}-e^{\gamma (\tau_\mathrm{p}-\tau_\mathrm{t})}\right]}{\gamma \tau_\mathrm{t}[(\gamma\tau_\mathrm{t}/2\pi)^2+1](e^{\gamma \tau_\mathrm{p}}-1)}\right].
\label{statecyclone}
\end{align}
This is the main result of this paper.
Since $R_{12}$ is proportional to $\sin(\phi_2-\phi_1)$, it vanishes when $\phi_1=\phi_2$.
In other words, the state variables $s_1$ and $s_2$ should be out-of-phase ($\phi_1 \neq \phi_2$) 
with respect to each other in order to exhibit a functionality.
This result is in accordance with the scallop theorem for a microswimmer~\cite{Golestanian08,Shapere89}.
Moreover, the nonreciprocality satisfies the symmetry property such that $R_{12}=-R_{21}$.

From Eq.~(\ref{statecyclone}), the asymptotic expressions of $R_{12}$ can be obtained as 
\begin{align}
R_{12}&\approx  \frac{d^2\gamma^2 \tau_\mathrm{p}^2\sin(\phi_2-\phi_1)}{4\pi[(\gamma\tau_\mathrm{p}/2\pi)^2+1]}~~~(\tau_\mathrm{t}/\tau_\mathrm{p}\approx 1),
\label{CA1}\\
R_{12}&\approx
\frac{3d^2\gamma^2 \tau_\mathrm{t}^2}{4\pi}\sin(\phi_2-\phi_1)~~~(\tau_\mathrm{t}/\tau_\mathrm{p}\ll1).
\label{CA2}
\end{align}
In Fig.~\ref{Fig:C}, we plot the dimensionless nonreciprocality $r_{12}$, i.e., $R_{12}$ scaled by the right 
hand side of Eq.~(\ref{CA1}), as a function of the ratio $\tau_\mathrm{t}/\tau_\mathrm{p}$ for 
different values of $\gamma \tau_\mathrm{p}$.
The dotted lines represent the asymptotic expression in Eq.~(\ref{CA2}).
From this plot, one can confirm the scaling behavior $R_{12} \sim (\gamma \tau_\mathrm{t})^2$ when 
$\tau_\mathrm{t}/\tau_\mathrm{p} \ll 1$. 
When $\tau_\mathrm{t}/\tau_\mathrm{p} \approx 1$, on the other hand, $r_{12}$ approaches unity as 
we see in Eq.~(\ref{CA1}).

\begin{figure*}[t]
\begin{center}
\includegraphics[scale=0.5]{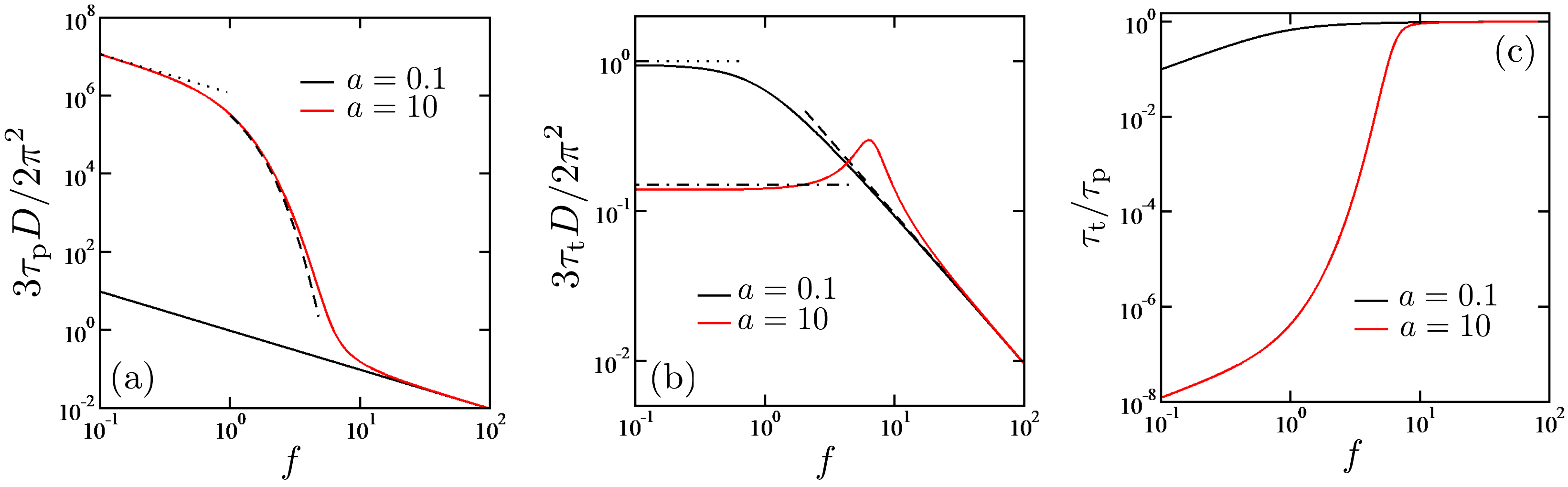}
\end{center}
\caption{
(a) The dimensionless mean first passage time $\tau_\mathrm{p}$ as a function of the dimensionless
nonequilibrium force $f$ for $a=0.1$ (black) and $10$ (red or gray).   
The plot for $a =0.1$ almost coincides with the asymptotic form in Eq.~(\ref{TpA1}).
The dotted and dashed lines are the asymptotic expressions in Eqs.~(\ref{TpA2}) and (\ref{TpA3}), respectively.
(b) The dimensionless mean first transition path time $\tau_\mathrm{t}$ as a function of the dimensionless
nonequilibrium force $f$ for $a=0.1$ (black) and $10$ (red or gray). 
The dotted and dashed lines are the asymptotic expressions in Eqs.~(\ref{TtA1}) and (\ref{TtA2}), respectively.
The dashed-dotted line is the asymptotic expression in Eq.~(\ref{TtA3}).
(c) The ratio $\tau_\mathrm{t}/\tau_\mathrm{p}$ as a function of the dimensionless
nonequilibrium force $f$ for $a=0.1$ and $10$.  
The asymptotic value of $\tau_\mathrm{t}/\tau_\mathrm{p}$ approaches unity for large $f$, 
while it strongly depends on $a$ for small $f$. 
}
\label{Fig:Tp}
\end{figure*}

\section{Two characteristic time scales}
\label{time}

As discussed in Sec.~\ref{state}, the dynamics of a catalytic chemical reaction is generally 
characterized by the mean first passage time $\tau_\mathrm{p}$ and the mean first transition 
path time $\tau_\mathrm{t}$.
According to the Kramers theory, $\tau_\mathrm{p}$ gives the time to overcome an energy 
barrier, and the inverse of it is a chemical reaction rate~\cite{Hanggi90}. 
While most of the first passage time is spent by the waiting time, the actual time required for 
a state transition can be much smaller.
Such a short time scale is characterized by  
$\tau_\mathrm{t}$~\cite{Laleman17,Kim15,Chaudhury10,Hummer04,Caraglio20}.
For a nucleic acid folding, it was estimated to be $\tau_\mathrm{t}\le10^{-5}$~s ~\cite{Neupane12,Chung13,Neupane17}.

In this section, we obtain the analytical expressions of $\tau_\mathrm{p}$ and $\tau_\mathrm{t}$ in 
terms of the potential parameters $A$ and $F$ in $G_{\rm r}$. 
Within the decoupling approximation, $K d^2 \ll A$, we consider the dynamics of $\theta$ by 
using Eq.~(\ref{Dt3}). 
Then one can obtain both $\tau_\mathrm{p}$ and $\tau_ \mathrm{t}$ for a general potential~\cite{Kim15}.

\subsection{Mean first passage time $\tau_\mathrm{p}$}

The first passage time $t_\mathrm{p}$ is a time for a reaction that started from the initial value 
$\theta_0$ reaches the final value $\theta_\mathrm{f}$ for the first time.
Notice that $\theta_0= 0$ and $\theta_\mathrm{f}=2\pi$ in our model.
Mathematically, this is equivalent to consider a Brownian motion of a particle in a semi-infinite system with an 
absorbing boundary condition at $\theta=\theta_\mathrm{f}$, and to measure the time until the particle is 
absorbed at $\theta=\theta_\mathrm{f}$ [see Fig.~\ref{Fig:TimeScale} (a)].
However, because the state $\theta=\theta_0$ can be visited multiple times, most of the first passage 
time is spent by a long waiting time.
Since $t_\mathrm{p}$ is a stochastic quantity and has a broad distribution, it is useful to consider its 
average value $\tau_\mathrm{p}$ called the mean first passage time.

A formal derivation of $\tau_\mathrm{p}$ is reviewed in Appendix~\ref{appMFPT}. 
For an arbitrary periodic function $G_\mathrm{p}$, $\tau_\mathrm{p}$ can be given by 
\begin{align}
\tau_\mathrm{p}&=\frac{1}{D(1-e^{-2\pi f})}\nonumber\\
&\times\int_{0}^{2\pi}dx\,\int_{0}^{2\pi}dy\,\exp[g_\mathrm{p}(x)-g_\mathrm{p}(x-y)-f y],
\label{Tpg}
\end{align}
where $g_\mathrm{p}=G_\mathrm{p}/(k_\mathrm{B}T)$ and $f =F/(k_\mathrm{B}T)$
are the dimensionless potential and nonequilibrium force, respectively.
Moreover, we have defined the diffusion constant $D=Mk_\mathrm{B}T$, where $M$ is 
the Onsager coefficient used in Eqs.~(\ref{Dt1}) and (\ref{Dt3}).
Note that $D$ has the dimension of inverse time in the present model.

The periodic potential $G_\mathrm{p}(\theta)$ in Eq.~(\ref{Gt}) should satisfy 
$G_\mathrm{p}(\theta+2\pi)=G_\mathrm{p}(\theta)$ and its energy barrier is $A$. 
Among various possibilities, the simplest form would be $G_\mathrm{p}(\theta)=-A\cos\theta$ 
that was used in our numerical simulation.
However, to perform the integral in Eq.~(\ref{Tpg}) analytically, we employ here the following 
linear functions: 
\begin{align}
G_\mathrm{p}(\theta)=\left\{
\begin{array}{ll}
A\left( \dfrac{2}{\pi}\theta-1 \right) &{\rm for}~~~0\le \theta<\pi \\
A\left( -\dfrac{2}{\pi}\theta+3 \right) &{\rm for}~~~\pi\le\theta<2\pi
\label{Gp}
\end{array}
\right.,
\end{align}
as depicted in Fig.~\ref{Fig:pot}(a).
With this periodic potential, one can analytically obtain $\tau_\mathrm{p}$ as
\begin{align}
\tau_\mathrm{p}&=\frac{2\pi^2}{D}\left[\frac{\pi f}{\pi^2 f^2-4 a^2}-\frac{8 a^2}{(\pi^2 f^2-4 a^2)^2}\frac{1+e^{-2\pi f}}
{1-e^{-2\pi f}}\right.\nonumber\\
&\left.+\frac{16 a^2\cosh(2 a)}{(\pi^2 f^2-4 a^2)^2}\frac{e^{-\pi f}}{1-e^{-2\pi f}}\right],
\label{Tp}
\end{align}
where $a=A/(k_\mathrm{B}T)$ is the dimensionless energy barrier.
Then the asymptotic expressions of $\tau_\mathrm{p}$ are given as follows:
\begin{align}
\tau_\mathrm{p}&\approx \frac{2\pi}{D f}=\frac{2\pi}{MF}~~~(a\ll1~{\rm or}~a \ll f),
\label{TpA1}\\
\tau_\mathrm{p}&\approx \frac{\pi e^{2 a }}{2 D a^2 f}=\frac{\pi (k_\mathrm B T)^2e^{2 A/k_\mathrm B T }}
{2 M A^2 F}~~~(a\gg1~{\rm and}~f\ll 1),
\label{TpA2}\\
\tau_\mathrm{p}&\approx \frac{\pi^2 e^{2 a}e^{-\pi f}}{D a^2}=\frac{\pi^2k_\mathrm B T e^{2 A/k_\mathrm B T}e^{-\pi F/k_\mathrm B T}}{M A^2}\nonumber\\
&~~~(a \gg f\gg1).
\label{TpA3}
\end{align}
Here we have recovered the dimension in the last expressions for the clarity sake.
Since Eq.~(\ref{TpA1}) does not depend on the temperature, thermal fluctuations are irrelevant in this limit. 
This is not the case for Eqs.~(\ref{TpA2}) and (\ref{TpA3}) which diverge when the temperature vanishes
due to the exponential factors.

In Fig.~\ref{Fig:Tp}(a), we plot the scaled $\tau_\mathrm{p}$ in Eq.~(\ref{Tp}) as a function of $f$ for $a=0.1$ and $10$.
For $a =0.1$ (black), the entire behavior is simply approximated by Eq.~(\ref{TpA1}). 
For $a=10$ (red or gray), on the other hand, we have plotted Eqs.~(\ref{TpA2}) (dotted line)
and (\ref{TpA3}) (dashed line) which are in good agreement with the full expression of $\tau_\mathrm{p}$.

\subsection{Mean first transition path time $\tau_\mathrm{t}$}

The first transition path time $t_\mathrm{t}$ is a time for a reaction that started from the initial 
value $\theta_0$ reaches the final value $\theta_\mathrm{f}$ without returning to $\theta_0$.
Mathematically, this is equivalent to consider a Brownian motion of a particle in a finite system with absorbing boundary conditions 
both at $\theta=\theta_0$ and $\theta=\theta_\mathrm{f}$, and to measure the time until the particle is absorbed 
at $\theta=\theta_\mathrm{f}$ [the black trajectory in Fig.~\ref{Fig:TimeScale} (b)].
When the particle is absorbed at $\theta=\theta_0$, such an event is not counted [the red (gray) trajectory in 
Fig.~\ref{Fig:TimeScale} (b)].
Since $t_\mathrm{t}$ is also a random quantity, we consider its average $\tau_\mathrm{t}$ called 
the mean first transition path time.

A formal derivation of $\tau_\mathrm{t}$ is explained in Appendix~\ref{appMFTT}, and the result is given by 
\begin{align}
\tau_\mathrm{t}&=\frac{1}{D}\left[\int_0^{2\pi}dw\,\exp[g_\mathrm{r}(w) ] \right]^{-1}\nonumber\\
&\times\int_{0}^{2\pi}dx\int_x^{2\pi}dy\int_0^{x}dz\,\exp[-g_\mathrm{r}(x)+g_\mathrm{r}(y)+g_\mathrm{r}(z)],
\label{Ttg}
\end{align}
where $g_\mathrm{r}=G_\mathrm{r}/(k_\mathrm{B}T)$. 
Using Eqs.~(\ref{Gt}) and (\ref{Gp}), we can analytically obtain $\tau_\mathrm{t}$ as 
\begin{align}
\tau_\mathrm{t}&=\frac{2\pi^2}{D[2 a(1-2e^{2a}e^{\pi f}+e^{2\pi f})+\pi f(e^{2\pi f}-1)]}\nonumber\\
&\times\frac{\Xi_0+\Xi_1(\pi f)+\Xi_2(\pi f)^2+\Xi_3(\pi f)^3+\Xi_4(\pi f)^4}
{(\pi^2 f^2-4 a^2)^2},
\label{Tt}
\end{align}
where
\begin{align}
\Xi_0&=-2(2 a)^4e^{2 a}e^{\pi f}\nonumber\\
&+(2 a)^3(e^{-2 a}e^{\pi f}+3e^{2 a}e^{\pi f}-2-2e^{2 \pi f}),\\
\Xi_1&=(2 a)^3(1-e^{2\pi f})+3(2 a)^2(1-e^{2\pi f}),\\
\Xi_2&=(2 a)^2(-1+2e^{2 a}e^{\pi f}-3e^{2\pi f})\nonumber\\
&+(2 a)(-e^{-2 a}e^{\pi f}+5e^{2 a}e^{\pi f}-2e^{2\pi f}-2),\\
\Xi_3&=2 a(e^{2\pi f}-1)-(e^{2\pi f}-1),\\
\Xi_4&=e^{2\pi f}+1.
\end{align}
Then the asymptotic expressions of $\tau_\mathrm{t}$ are given as follows:
\begin{align}
\tau_\mathrm{t}&\approx \frac{2\pi^2}{3D}=\frac{2\pi^2}{3Mk_\mathrm BT}~~~(a\ll1~{\rm and}~f\ll 1),
\label{TtA1}\\
\tau_\mathrm{t}&\approx \frac{2\pi}{D f}=\frac{2\pi}{MF}~~~(f \gg1~{\rm and}~a \ll f),
\label{TtA2}\\
\tau_\mathrm{t}&\approx \frac{\pi^2}{D a}=\frac{\pi^2}{MA}~~~(a \gg1~{\rm and}~a \gg f).
\label{TtA3}
\end{align}
In the limit of Eq.~(\ref{TtA1}), the transition process is dominated by thermal fluctuations.
On the other hand, Eqs.~(\ref{TtA2}) and (\ref{TtA3}) are independent of the temperature, and hence the 
transitions occur deterministically.
The scaling relation $\tau_\mathrm{t} \sim 1/a$ in Eq.~(\ref{TtA3}) was also obtained before for a quadratic 
potential~\cite{Chung09}.

In Fig.~\ref{Fig:Tp}(b), we plot the scaled $\tau_\mathrm{t}$ as a function of $f$ for $a=0.1$ and $10$.
For $a=0.1$ (black), $\tau_\mathrm{t}$ is constant for $f<1$ and it decreases for $f>1$. 
This behavior is in accordance with the asymptotic expressions in Eq.~(\ref{TtA1}) (dotted line)
and Eq.~(\ref{TtA2}) (dashed line).
For $a=10$ (red or gray), on the other hand, $\tau_\mathrm{t}$ takes a maximum value around $f \approx a$.
The dashed-dotted line is the asymptotic expression in Eq.~(\ref{TtA3}).

In Fig.~\ref{Fig:Tp}(c), we plot the ratio $\tau_\mathrm{t}/\tau_\mathrm{p}$ as a function of $f$ for 
$a=0.1$ and $10$.
For $a=0.1$ (black), a power law dependence is seen for $f<1$, and $\tau_\mathrm{t}/\tau_\mathrm{p}$ 
approaches unity for $f>1$.
For $a=10$ (red or gray), on the other hand, $\tau_\mathrm{t}/\tau_\mathrm{p}$ is vanishingly small for $f < 1$, 
and it grows exponentially for $1< f < a$.

It is worth mentioning here the characteristic difference between $\tau_\mathrm{p}$ and $\tau_\mathrm{t}$.
In the limit of $f \to 0$, $\tau_\mathrm{p}$ diverges 
[see Eqs.~(\ref{TpA1}) and (\ref{TpA2})] while $\tau_\mathrm{t}$ remains finite
[see Eqs.~(\ref{TtA1}) and (\ref{TtA2})].
This is because a nonequilibrium driving force is always required for the net chemical reaction with a 
finite $\tau_\mathrm{p}$. 
On the other hand, $\tau_\mathrm{t}$ can be evaluated even in the equilibrium situation.

\section{Summary and discussion}
\label{diss}

In this paper, we have discussed cyclic state transitions of a micromachine driven by a catalytic 
chemical reaction.
We have proposed a minimum model of a generic micromachine and calculated the nonreciprocality in 
Eq.~(\ref{Cyclon}) to quantify the functionality.
Our model uses the reaction variable $\theta$ and the state variables $s_i$ which are coupled 
to each other through the mechano-chemical coupling mechanism.
The tilted periodic potential $G_{\rm r}(\theta)$ for catalytic reaction is characterized by the 
energy barrier $A$ and the nonequilibrium force $F$.
Importantly, the state variables $s_i$ are required to change periodically in time for a catalytic reaction.

To investigate our model analytically, we have employed the decoupling approximation for
the Onsager's equations. 
Furthermore, we have assumed that the reaction variable $\theta$ obeys a deterministic stepwise 
dynamics characterized by the mean first passage time $\tau_\mathrm{p}$ and the mean first 
transition path time $\tau_\mathrm{t}$.
Under these assumptions, we have first obtained the time dependencies of the state variables
$s_{1}$ and $s_{2}$ in terms of $\tau_\mathrm{p}$ and $\tau_\mathrm{t}$ 
[see Eqs.~(\ref{Ss1}) and (\ref{Ss2})].
We find that the magnitudes of $s_{1}$ and $s_{2}$ become smaller when  
$\tau_\mathrm{t}/\tau_\mathrm{p}$ or $\gamma \tau_\mathrm{p}$ is decreased.
Then we have analytically obtained the nonreciprocality $R_{12}$ within the decoupling 
approximation [see Eq.~(\ref{statecyclone})].
One of the important results is the asymptotic scaling relation  
$R_{12}\sim (\gamma \tau_\mathrm{t})^2$ for $\tau_\mathrm{t}/\tau_\mathrm{p}\ll1$
[see Eq.~(\ref{CA2})].
Using Eq.~(\ref{Dt3}) in the small coupling limit, we have further obtained $\tau_\mathrm{p}$ 
[see Eq.~(\ref{Tp})] and $\tau_\mathrm{t}$ [see Eq.~(\ref{Tt})] in terms of the potential 
parameters $A$ and $F$.

So far, the nonreciprocality $R_{12}$ has been obtained in terms of $\tau_\mathrm{p}$ 
and $\tau_\mathrm{t}$, whereas they are further expressed in terms of $A$ and $F$.
For a realistic chemical reaction such as ATP hydrolysis, $\tau_\mathrm{t}/\tau_\mathrm{p}$ 
is typically small, and we expect that $R_{12}$ can be well approximated by Eq.~(\ref{CA2}). 
On the other hand, the limit of Eq.~(\ref{TtA3}) is appropriate for $\tau_\mathrm{t}$ when 
$a=A/(k_\mathrm{B}T)$ is large enough.
Using the corresponding asymptotic expressions, the relevant scaling for the nonreciprocality 
$R_{12}$ turns out to be 
\begin{align}
R_{12} \sim \left(\frac{d\gamma}{Da} \right)^{2}\sin(\phi_2-\phi_1) 
 \sim \left( \frac{d\mu K}{MA} \right)^{2}\sin(\phi_2-\phi_1).
\label{Cest}
\end{align}
In particular, the relation $R_{12}\sim 1/A^2$ implies that the higher the energy barrier is, 
the smaller the nonreciprocality becomes.
This scaling relation is another important result of the present model.

Next we discuss typical values of the model parameters. 
From the experiment measuring the enhanced diffusion of a motor protein, the energy barrier 
of ATP was estimated to be $A\sim 10\,k_\mathrm{B}T$~\cite{Hayashi15}.
When a single ATP molecule is converted into ADP, the produced energy is roughly $20\,k_\mathrm{B}T$~\cite{Toyabe10}.
Then we estimate the nonequilibrium chemical force as $F\sim20/(2\pi)\,k_\mathrm{B}T \sim 3\,k_\mathrm{B}T$
(notice again that the dimension of $F$ is energy).
Hence both $A/k_\mathrm{B}T  >1$ ($a>1$) and $A > F$ ($a>f$) are satisfied.
Moreover, one can estimate from Eqs.~(\ref{TpA2}) and (\ref{TtA3}) that $\tau_\mathrm{t}/\tau_\mathrm{p}\sim10^{-6}\ll1$, 
which justifies the assumption in Eq.~(\ref{Cest}).

Recent experiments reported the diffusion enhancement of enzymes due to catalytic chemical reactions~\cite{Muddana10,Jee18}.
When a self-propelled particle undergoes a rotational diffusion, its translational diffusion coefficient increases by 
$\Delta D=V^2\tau_{\rm rot}$, where $V$ is the propulsion velocity and $\tau_{\rm rot}$ is the rotational diffusion 
time~\cite{Jiang10}. 
Since the propulsion velocity is proportional to the nonreciprocality, $V\sim R$, the diffusion enhancement 
can be estimated as $\Delta D\sim R^2 \tau_{\rm rot} \sim \tau_{\rm rot}/A^4$. 
So far, the relation between the energy barrier and the functionality of a micromachine has not yet been investigated.
We predict that the change in the energy barrier can be reflected in the diffusion enhancement of 
enzymes.

In the present study, we have mainly discussed the case when there are only two degrees of freedom
($s_{1}$ and $s_{2}$) of a micromachine.
Although this is a minimum and sufficient number to discuss the nonreciprocality $R_{12}$, one needs to take 
into account a large number of state variables to describe the dynamics of realistic enzymes. 
As mentioned in Introduction, the total functionality of a micromachine can be expressed by the weighted sum
$\sum_{i,j}W_{ij}R_{ij}$, where $W_{ij}$ is the weighting tensor that depends on the properties of a micromachine.
Hence it is necessary to evaluate the nonreciprocalities $R_{ij}$ for all the binary combinations of the state variables. 
Although the estimation of the weight tensor $W_{ij}$ is beyond the scope of this work, such a study will 
be important in the future.

In the present work, the nonreciprocality $R_{12}$ has been obtained only in the weak coupling limit, $Kd^2 \ll A$. 
The investigation of the opposite strong coupling limit is also left as a future work such as performing more extended 
numerical simulations. 
It would be also interesting to see the case when the off-diagonal elements of the Onsager coefficient 
is nonzero, i.e., $\mu_{ij} \neq 0$ for $i \neq j$.

\acknowledgements

K.Y.\ acknowledges support by a Grant-in-Aid for JSPS Fellows (Grant No.\ 18J21231) from the Japan Society 
for the Promotion of Science (JSPS). 
S.K.\ acknowledges support by a Grant-in-Aid for Scientific Research (C) (Grant No.\ 18K03567 and
Grant No.\ 19K03765) from the JSPS, 
and support by a Grant-in-Aid for Scientific Research on Innovative Areas
``Information Physics of Living Matters'' (Grant No.\ 20H05538) from the Ministry of Education, Culture, 
Sports, Science and Technology of Japan.

\appendix
\section{Derivation of Eqs.~(\ref{Ss1}) and (\ref{Ss2})}
\label{appDs}

In this Appendix, we show the derivation of Eqs.~(\ref{Ss1}) and (\ref{Ss2}).
In the absence of the noise terms, Eqs.~(\ref{Ds2a}) and (\ref{Ds2b}) can be formally solved as 
\begin{align}
\delta_i(t)=-e^{-\gamma t}\int_{-\infty}^t dt'\,e^{\gamma t'}d\cos[\theta(t')+\phi_i]\dot\theta(t'),
\label{solapp}
\end{align}
where we have assumed $0 \le t \le \tau_\mathrm{p}$ and ignored a term that depends on the initial condition.
Using Eq.~(\ref{Asst2}) for $\theta$ and the condition $\theta(t+\tau_\mathrm{p})=\theta(t)+2\pi$, we obtain 
\begin{align}
\delta_i(t)&=-\frac{2\pi}{\tau_\mathrm t} d e^{-\gamma t}\left [\int_{0}^t dt' \,e^{\gamma t'}\cos\left(\frac{2\pi t'}{\tau_\mathrm t}+\phi_i\right)\Theta(\tau_\mathrm{t}-t')\right.\nonumber\\
&\left.-\sum_{n=1 }^{\infty}\int_{-n\tau_\mathrm{p}}^{-n\tau_\mathrm{p}+\tau_\mathrm{t}} dt'\,e^{\gamma t'}\cos\left[ \frac{2\pi}{\tau_\mathrm t} (t'+n\tau_\mathrm{p})+\phi_i\right] \right],
\end{align}
where $\Theta(t)$ is the Heaviside step function. 
Changing the variable to $t''=t'+n\tau_\mathrm{p}$ in the second integral, we obtain
\begin{align}
\delta_i(t)&=-\frac{2\pi}{\tau_\mathrm t} d e^{-\gamma t}\left [\int_{0}^t dt' \,e^{\gamma t'}\cos\left(\frac{2\pi t'}{\tau_\mathrm t}+\phi_i\right)\Theta(\tau_\mathrm{t}-t')\right.\nonumber\\
&\left.-\sum_{n=1 }^{\infty}e^{-\gamma n\tau_\mathrm{p}}\int_{0}^{\tau_\mathrm{t}} dt''\,e^{\gamma t''}\cos \left(\frac{2\pi t''}{\tau_\mathrm t}+\phi_i\right)\right],
\end{align}
which results in Eqs.~(\ref{Ss1}) and (\ref{Ss2}) after the integration.

\section{Derivation of Eq.~(\ref{Tpg})}
\label{appMFPT}

In this Appendix, we show the derivation of the mean first passage time $\tau_\mathrm{p}$ in 
Eq.~(\ref{Tpg})~\cite{Hanggi90,Reimann01,Hayashi04}.
For this purpose, we consider a conditional probability distribution $P(\theta,t|\theta_0)$ for which 
$\theta(0)=\theta_0$ is imposed as the initial condition.
Then $P(\theta,t|\theta_0)$ satisfies the following Fokker-Planck equation:
\begin{align}
&\partial_tP(\theta,t|\theta_0)=\mathcal L(\theta)P(\theta,t|\theta_0), \\
&\mathcal L(\theta)=D\partial_\theta \exp[-g_\mathrm{r}(\theta)] \partial_\theta 
\exp[g_\mathrm{r}(\theta)].
\end{align}
Similarly, $P(\theta,t|\theta_0)$ also satisfies the following backward Fokker-Planck equation:
\begin{align}
&\partial_tP(\theta,t|\theta_0)=\mathcal L^\dag(\theta_0)P(\theta,t|\theta_0), 
\label{backward} \\
&\mathcal L^\dag(\theta_0)=D\exp[g_\mathrm{r}(\theta_0)]\partial_{\theta_0} 
\exp[-g_\mathrm{r}(\theta_0)]\partial_{\theta_0}.
\end{align}

We employ the reflective boundary condition at $\theta\to-\infty$ and the absorbing boundary condition at $\theta=2\pi$.
Then the total probability distribution decays due to the latter boundary condition. 
Here we introduce the survival probability defined as 
\begin{align}
S(t,\theta_0)=\int_{-\infty}^{2\pi}d\theta\, P(\theta,t|\theta_0).
\label{S}
\end{align}
Then the distribution function of the first passage time is given by 
\begin{align}
K_\mathrm{p}(t,\theta_0)=-\frac{dS}{dt}.
\label{Kp}
\end{align}
From the condition $S(0)=1$, the following normalization condition holds 
\begin{align}
\int_0^\infty dt\,K_\mathrm{p}(t,\theta_0)=1.
\label{KpN}
\end{align}
The mean first passage time $\tau_\mathrm{p}$ is defined as the first moment of the distribution function
\begin{align}
\tau_\mathrm{p}(\theta_0)=\int_0^{\infty}dt\, tK_\mathrm{p}(t,\theta_0).
\label{Tpdef}
\end{align}

Next, one can show from  Eqs.~(\ref{backward}), (\ref{S}), (\ref{Kp}), and (\ref{Tpdef}) that 
\begin{align}
-1=\mathcal L^\dag(\theta_0)\tau_\mathrm{p}(\theta_0), 
\end{align}
where we have used the conditions $\lim_{t\to0}tK_\mathrm{p}(t)=0$ and $\lim_{t\to\infty}tK_\mathrm{p}(t)=0$.
Using the reflective boundary condition ($\partial_{\theta_0}\tau_\mathrm{p}=0$ at $\theta_0\to-\infty$)
and the absorbing boundary conditions ($\tau_\mathrm{p}=0$ at $\theta_0=2\pi$), one can solve the 
above equation to obtain~\cite{Goel74},
\begin{align}
\tau_\mathrm{p}(\theta_0)=\frac{1}{D}\int_{\theta_0}^{2\pi}dx\,\exp[g_\mathrm{r}(x)]\int_{-\infty}^{x}dY\,\exp[-g_\mathrm{r}(Y)].
\end{align}
Setting $\theta_0=0$ and using Eq.~(\ref{Gt}), we get 
\begin{align}
\tau_\mathrm{p}=\frac{1}{D}\int_{0}^{2\pi}dx\int_{0}^{\infty}dy'\,
\exp[g_\mathrm{p}(x)-g_\mathrm{p}(x-y')-fy'],
\end{align}
where $y'=x-Y$. 
Because the periodicity of $G_\mathrm{p}$ is $2\pi$, we obtain 
\begin{align}
\tau_\mathrm{p}&=\frac{1}{D}\sum_{n=0}^\infty e^{-2n\pi F}\nonumber\\
&\times \int_{0}^{2\pi}dx\int_{0}^{2\pi}dy\,
\exp[g_\mathrm{p}(x)-g_\mathrm{p}(x-y)-fy],
\end{align}
where $y=y'-2n\pi$. 
Since $F>0$, we can easily evaluate the infinite series and obtain Eq.~(\ref{Tpg}).

\section{Derivation of Eq.~(\ref{Ttg})}
\label{appMFTT}

In this Appendix, we show the derivation of the mean first transition path time $\tau_\mathrm{t}$ in 
Eq.~(\ref{Ttg})~\cite{Kim15,Chaudhury10,Hummer04}.
In this case, the absorbing boundary condition is imposed both at $\theta=0$ and $\theta=2\pi$.
Hence, unlike $\tau_\mathrm{p}$, the probability is absorbed from both of the boundaries, although 
the probability of being absorbed at $\theta=2\pi$ determines $\tau_\mathrm{t}$.

The distribution function of the first transition path time is given by
\begin{align}
K_\mathrm{t}(t,\theta_0)=\frac{1}{N} J(\theta=2\pi,t|\theta_0),
\label{Kt}
\end{align}
where $J(\theta,t|\theta_0)$ is a probability flux 
\begin{align}
J(\theta,t|\theta_0)=-D \exp[-g_\mathrm{r}(\theta)] \frac{\partial}{\partial \theta}
\left[\exp[g_\mathrm{r}(\theta)]P(\theta,t|\theta_0) \right].
\label{J}
\end{align}
In Eq.~(\ref{Kt}), $N$ is the normalization constant that is fixed by the condition
\begin{align}
\int_0^\infty dt\,K_\mathrm{t}(t,\theta_0)=1.
\label{KtN}
\end{align}
The mean first transition path time $\tau_\mathrm{t}$ is defined as the first moment of the distribution function
\begin{align}
\tau_\mathrm{t}(\theta_0)=\int_0^{\infty}dt\, tK_\mathrm{t}(t,\theta_0).
\label{Ttdef}
\end{align}

The backward Fokker-Planck equation for the probability flux $J$ is given by
\begin{align}
\partial_tJ(\theta,t|\theta_0)=\mathcal L^\dag(\theta_0)J(\theta,t|\theta_0).
\label{Jeq}
\end{align}
From Eqs.~(\ref{Kt}), (\ref{KtN}), and (\ref{Jeq}), one can show that 
\begin{align}
0=\mathcal L^\dag(\theta_0)N(\theta_0),
\end{align}
where we have used the conditions $J(2\pi,0|\theta_0)=0$ and $J(2\pi,\infty|\theta_0)=0$.
Solving this equation with the boundary conditions $N(0)=0$ and $N(2\pi)=1$, we obtain
\begin{align}
N(\theta_0)=\int_{0}^{\theta_0}dy_0\,\exp[g_\mathrm{r}(y_0)]
\left[ \int_{0}^{2\pi}dy_0\,\exp[g_\mathrm{r}(y_0)] \right]^{-1}.
\end{align}

From Eqs.~(\ref{Kt}), (\ref{Ttdef}), and (\ref{Jeq}), we obtain 
\begin{align}
-N(\theta_0)=\mathcal L^\dag(\theta_0)\psi(\theta_0), 
\end{align}
where $\psi(\theta_0)=\tau_\mathrm{t}(\theta_0)N(\theta_0)$ and we have used the conditions 
$\lim_{t\to0}tK_\mathrm{t}(t,\theta_0)=0$ and $\lim_{t\to\infty}tK_\mathrm{t}(t,\theta_0)=0$.
With the use of the absorbing boundary conditions $\psi(0)=0$ and $\psi(2\pi)=0$, we 
can solve the above equation to obtain
\begin{align}
\psi(\theta_0)& =\frac{1}{D}\left[\int_{0}^{2\pi}dw\,\exp[g_\mathrm{r}(w)]\right]^{-1}
\nonumber \\
& \times \left[(1-N(\theta_0))\int_{0}^{\theta_0}dx\,\exp[-g_\mathrm{r} (x)]N^2(x)\right.\nonumber\\
&\left.+N(\theta_0)\int_{\theta_0}^{2\pi}dx\,\exp[-g_\mathrm{r}(x)](1-N(x))N(x)\right].
\label{psif}
\end{align}
Since $\tau_\mathrm{t}=\lim_{\theta_0\to 0}[\psi(\theta_0)/N(\theta_0)]$, only the second term remains and we  
obtain Eq.~(\ref{Ttg}).



\end{document}